\documentclass[aoas,nameyear,rotating,dvips]{arximspdf}
\usepackage{dcolumn,lettersp}
\usepackage{graphicx}

\doi{10.1214/09-AOAS300}
\volume{4}
\issue{2}
\pubyear{2010}
\firstpage{988}
\lastpage{1013}

\makeatletter

\newcolumntype{d}[1]{D{.}{.}{#1}}
\newcommand{\ph}{\phantom{00}}
\makeatother

\begin{document}
\begin{frontmatter}

\title{Bayesian meta-analysis for identifying periodically expressed
genes in fission yeast cell cycle\protect\thanksref{TZ}}
\runtitle{Bayesian meta-analysis of cell cycle gene expression}
\thankstext{TZ}{Supported in part by the
NIH Grant R01GM078990 and the NSF Grant DMS-07-06989.}

\begin{aug}
\author[A]{\fnms{Xiaodan} \snm{Fan}\ead[label=e1]{xfan@stat.harvard.edu}\ead[label=e4]{xfan@sta.cuhk.edu.hk}},
\author[B]{\fnms{Saumyadipta} \snm{Pyne}\ead[label=e2]{spyne@broad.mit.edu}}
\and
\author[C]{\fnms{Jun S.} \snm{Liu} \ead[label=e3]{jliu@stat.harvard.edu}\corref{}}
\runauthor{X. Fan, S. Pyne and S. Liu}
\affiliation{Harvard University and Chinese University of Hong Kong, Broad Institute\break and Harvard University}
\address[A]{X. Fan\\
Department of Statistics\\
Harvard University\\
One Oxford Street\\
Cambridge, Massachusetts 02138\\
USA\\
and\\
Department of Statistics\\
Chinese University of Hong Kong\\
Shatin, N.T.\\
Hong Kong\\
\printead{e1}\\
\phantom{E-mail: }\printead*{e4}}

\address[B]{S. Pyne\\
Broad Institute of MIT and Harvard\\
7 Cambridge Center\\
Cambridge, Massachusetts 02142\\
USA\\
\printead{e2}}

\address[C]{J. S. Liu\\
Department of Statistics\\
Harvard University\\
One Oxford Street\\
Cambridge, Massachusetts 02138\\
USA\\
\printead{e3}}

\end{aug}

% HISTORY:
\received{\smonth{1} \syear{2009}}
\revised{\smonth{9} \syear{2009}}

% ABSTRACT
\begin{abstract}
The effort to identify genes with periodic expression during the
cell cycle from genome-wide microarray time series data has been
ongoing for a decade. However, the lack of rigorous modeling of
periodic expression as well as the lack of a comprehensive model
for integrating information across genes and experiments has
impaired the effort for the accurate identification of
periodically expressed genes. To address the problem, we introduce
a Bayesian model to integrate multiple independent microarray data
sets from three recent genome-wide cell cycle studies on  fission
yeast. A hierarchical model was used for data integration. In
order to facilitate an efficient Monte Carlo sampling from the
joint posterior distribution, we develop a novel
Metropolis--Hastings group move. A surprising finding from our
integrated analysis is that more than 40\% of the genes in
fission yeast are significantly periodically expressed, greatly
enhancing the reported  10--15\% of the genes in the current
literature. It calls for a reconsideration of the periodically
expressed gene detection problem.
\end{abstract}

% KEYWORDS

\end{frontmatter}

%s1 ###
\section{Introduction}

The cell division cycle is the concerted sequence of process\-es by
which a cell duplicates its DNA and divides into two daughter
cells. Many genes are expressed periodically at a specific stage
during the cell cycle when they peak and trough over a certain
time range. They are termed as ``cell cycle-regulated genes.''
Here, in the context of mRNA expression studies, we call these
``Periodically Expressed (PE) genes.'' In contrast, other genes
are called ``APeriodically Expressed (APE) genes.'' Identification
of PE genes is both of theoretical importance because of the need
to understand the different mechanisms underlying these genes'
involvements in the cell cycle processes, and of practical
importance due to the biological links between cell cycle control
and many diseases such as cancer [\citet{refSherr96Science}; \citet{refWhitfieldBotstein02MolBiolCell}; \citet{refBarJosephSimon08PNAS}].

With the help of the microarray techniques and various cell phase
synchronization methods (synchronizing the progression of cells
through the stages of cell cycle), researchers have conducted
genome-wide time series expression analyses on synchronized cells
for various species ranging from fungi to plant to human
[\citet{refChoDavis98MolCell}; \citet{refSpellmanFutcher98MolBiolCell}; \citet{refLaubShapiro00Science};
\citet{refIshidaNevins01MolCellBiol}; \citet{refMengesMurray02JBiolChem};
\citet{refWhitfieldBotstein02MolBiolCell}; \citet{refRusticiBahler04NatGenet};
\citet{refOlivaLeatherwood05PLoSBiol};
\citet{refPengLiu05MolBiolCell};
\citet{refBarJosephSimon08PNAS}].
Several strategies for identifying PE genes on these data have
been developed, such as the fitting of a
sinusoidal function [\citet{refSpellmanFutcher98MolBiolCell}],
clustering techniques [\citet{refEisenBotstein98PNAS}; \citet{refWhitfieldBotstein02MolBiolCell}], the single-pulse model
[\citet{refZhaoBreeden01PNAS}], the partial least squares
regression approach [\citet{refJohanssonBerglund03Bioinfo}], the
average periodogram [\citet{refWichertStrimmer04Bioinfo}], the
linear combination of cubic $B$-spline basis
[\citet{refLuanLi04Bioinfo}], the random-periods model
[\citet{refLiuWeinberg04PNAS}], the least square fitting for the
periodic-normal mixture model [\citet{refLuLiu04NAR}], the
Fourier score combined with $p$-value of regulation
[\citet{refLichtenbergBrunak05Bioinfo}], the robust spectral
estimator combined with $g$-statistic
[\citet{refAhdesmakiYliHarja05BMCBioinfo}] and the up-down
signature method [\citet{refWillbrandFink05Bioinfo}].
\citet{refZhouBreeden05working} applied a Bayesian approach
for single experiment data by fixing the period at pre-estimated
value. Most of these methods use a set of known PE genes to
estimate the cell cycle period prior to testing the periodicity
for other genes.

While the previous efforts have often reported positively about
the presence of the periodic signal in these gene expression data,
doubts were raised as to whether such periodic gene regulation was
reproducible [\citet{refSheddenCooper02PNAS}; \citet{refWichertStrimmer04Bioinfo}] and, by extension, about the
identity and count of PE genes discovered by subsequent analyses.
One prevalent reason for skepticism is the reliance of many of
the studies on ad hoc thresholds to classify genes as PE or
otherwise. For example, \citet{refChoDavis98MolCell} detected
the PE genes by visual inspection;
\citet{refSpellmanFutcher98MolBiolCell} designed a cutoff
value based on prior biological knowledge. Another possible reason
is that the commonly assumed white noise background model for time
series might be too unrealistic to allow correct inference about
the identity and count of PE genes
[\citet{refFutschikHerzel08Bioinfo}]. Furthermore, all previous
approaches were designed for analyzing single time series per
gene, which did not allow for an efficient combination of data from
multiple experiments and therefore lacked the power to identify a
large fraction of all PE genes. Recently
\citet{refTsiporkovaBoeva08Bioinfo} proposed a procedure to
combine multi-experiment data based on a dynamic time warping
alignment technique, which is potentially useful for analyzing
multiple cell cycle data sets if combined with a periodicity
detection algorithm. However, the procedure requires each time point
within a
time series to be aligned to a time point within the other time
series, which is not always appropriate when the lengths of cell cycle
period, the sampled time ranges and the sampling frequencies
are all different between experiments.

Recently, three independent studies
[\citet{refRusticiBahler04NatGenet}; \citet{refOlivaLeatherwood05PLoSBiol}; \citet{refPengLiu05MolBiolCell}]
conducted elutriation and cdc25 block-release synchronization
experiments to measure genome-wide expression in the fission yeast
(\textit{Schizosaccharomyces pombe}) cell cycle. The results from
these three studies also showed discrepancies with regard to the
identity and count of PE genes. They reported 407, 747 and 750 PE
genes, respectively, with only 176 genes being common to all three
lists. However, the availability of 10 genome-wide experiments
produced by these three different labs has made the fission yeast
currently the organism with the largest cell cycle transcriptome
data, which provides us an opportunity to obtain a better
understanding of the cell cycle.
\citet{refMargueratBahler06Yeast} combined the ten data sets
from the three studies by multiplying $p$-values for gene regulation
and periodicity from each experiment. They concluded that no more
than about 500 PE genes can be reliably identified from the
combined data. While observing that well over 1000 fission yeast
genes could be periodically expressed and that each study had
detected a different subset of these, they attributed the
discrepancy to inconsistent gene naming, the use of different data
analysis methods and the use of arbitrary thresholds.

We investigated the PE gene identification problem by employing a
Bayesian approach to provide (1) a more realistic and
comprehensive model for the cell cycle time series data, and (2)
an efficient and rigorous way to combine data from multiple
experiments. A hierarchical model together with MCMC computation
is used to integrate different sources of variation and
correlation into a single coherent probabilistic framework. We
applied this approach to integrate the ten genome-wide time series
data sets. A striking finding from our analysis is that more than
2000 genes are significantly periodically expressed. This number
greatly enhances the count of possible cell cycle regulated genes
in the current literature. Most interestingly, our finding can be
visualized clearly from Figure~\ref{All_Heatmap}, which merely
displays the \textit{original} data, but with the genes ordered
according to our inferred periodicity strength and peaking phase.

%s2 ###
\section{Materials and methods}

In Section~\ref{section:data} we describe the cell cycle gene
expression data. In Section~\ref{section:model} we outline our
parametric model for cell cycle gene expression. The Bayesian
computation of the model is described in
Section~\ref{section:identifiability} and
Section~\ref{section:computation}. In
Section~\ref{section:strategy} we present our strategies for
distinguishing PE genes from APE genes based on the model fitting
results.

%s2.1 ###
\subsection{Microarray time series data}\label{section:data}

%t1 ###
\begin{sidewaystable}
\tablewidth=\textwidth
 \caption{Summary of the ten experiments for the fission yeast cell cycle}\label{DataOverviewTable}
\begin{tabular*}{\textwidth}{@{\extracolsep{\fill}}lcccccccccc@{}}
\hline
\textbf{Data set name}
&\multicolumn{5}{c}{\textbf{Rustici et al.}}
&\multicolumn{2}{c}{\textbf{Peng et al.}}
&\multicolumn{3}{c@{}}{\textbf{Oliva et al.}}
\\[-6pt]
&\multicolumn{5}{c}{\hrulefill}
&\multicolumn{2}{c}{\hrulefill}
&\multicolumn{3}{c@{}}{\hrulefill}\\
\textbf{Microarray type}
&\multicolumn{5}{c}{\textbf{spotted PCR array}}
&\multicolumn{2}{c}{\textbf{spotted oligo array}}
&\multicolumn{3}{c@{}}{\textbf{spotted PCR array}}
\\[-6pt]
&\multicolumn{5}{c}{\hrulefill}
&\multicolumn{2}{c}{\hrulefill}
&\multicolumn{3}{c@{}}{\hrulefill}\\
\textbf{Synchronization technique}
&\multicolumn{3}{c}{\textbf{elutriation}}
&\multicolumn{2}{c}{\textbf{cdc25}}
&\multicolumn{1}{c}{\textbf{elutriation}}
&\multicolumn{1}{c}{\textbf{cdc25}}
&\multicolumn{2}{c}{\textbf{elutriation}}
&\multicolumn{1}{c@{}}{\textbf{cdc25}}\\[-6pt]
&\multicolumn{3}{c}{\hrulefill}
&\multicolumn{2}{c}{\hrulefill}
&\multicolumn{1}{c}{\hrulefill}
&\multicolumn{1}{c}{\hrulefill}
&\multicolumn{2}{c}{\hrulefill}
&\multicolumn{1}{c@{}}{\hrulefill}
\\
\textbf{Experiment name}  &  \textbf{Exp1}  &  \textbf{Exp2}  &  \textbf{Exp3}
&  \textbf{Exp4}  &  \textbf{Exp5}  &  \textbf{Exp6}  &  \textbf{Exp7}  &  \textbf{Exp8}  &  \textbf{Exp9}  &  \textbf{Exp10}\\
\hline
Number of covered gene          &  4113  &  3921  &  4176  &  4281  &  4173  &  4263  &  4571  &  4543  &  4400  &  4727\\
Number of time point ($S_e$)    &  \ph20  & \ph 20  & \ph 20  &  \ph19  & \ph 18  &  \ph33  &  \ph38  &  \ph33  &  \ph50  &  \ph52\\
Time point frequency (min)      &  \ph15  & \ph 15  & \ph 15  &  \ph15  &  \ph15  & \ph 10  &  \ph10  &  15--21  &  2--10  &  10--15\\\hline
\end{tabular*}
\tabnotetext[]{tz}{\textit{Note}:\vspace*{-5pt}
\begin{longlist}[1.]
\item[1.] The data set Rustici et al. is downloaded from
\url{http://www.sanger.ac.uk/PostGenomics/S\_pombe/projects/cellcycle/}.
Peng et al. is downloaded from
\url{http://giscompute.gis.a-star.edu.sg/\textasciitilde gisljh/CDC/CDC\_dnld\_data.html}.
Oliva et al. is downloaded from
\url{http://publications.redgreengene.com/oliva\_plos\_2005/}.

\item[2.] The downloaded data set Rustici et al. has been
normalized on an array-by-array basis using an in-house
normalization script, which performs three steps: masking bad
spots, filtering lower quality spots, applying local window-based
normalization. Peng et al. has filtered low intensity
features (2-fold less than the background) and done LOWESS
normalization within array. Oliva et al. has been
normalized within array by the GenePix Pro software with
default setting.
\item[3.] Elutriation experiments are done to wild-type fission yeast,
where samples of uniformly sized cells are obtained. Because cell
size is correlated with cell cycle stage, these cells are
synchronized with respect to their position in the cycle. Cdc25
block-and-release experiments are done to the fission yeast strain
carrying the temperature-sensitive cdc25-22 mutant gene, where
cells are initially synchronized by blocking them at some
particular cell cycle stage, then releasing them from the block
 and taking samples at different times.
 \end{longlist}}
\end{sidewaystable}

We obtained the normalized gene expression data for ten
genome-wide experiments by three cell cycle microarray studies
[\citet{refRusticiBahler04NatGenet}; \citet{refOlivaLeatherwood05PLoSBiol};
\citet{refPengLiu05MolBiolCell}]
from the websites listed in Table~\ref{DataOverviewTable}. For
each experiment, a culture of cells is grown and synchronized. A
set of microarrays is used to measure gene expressions at
selected time points (possibly with technical replication of the
microarray). All values were converted to log-ratios with base 2.
To make the log-ratios comparable across arrays, we transformed
the values for every array separately to set the median log-ratio
of each array to zero. Log-ratios from technical replicates, if
present, were averaged. Time series with more than 25 percent
missing entries were omitted. We unified gene names across the
studies based on GeneDB database entries
[\citet{refHertzFowlerBarrell04NAR}]. The genes without a
consistent nomenclature were excluded.

\begin{sidewaysfigure}
\vspace*{6pt}

\includegraphics{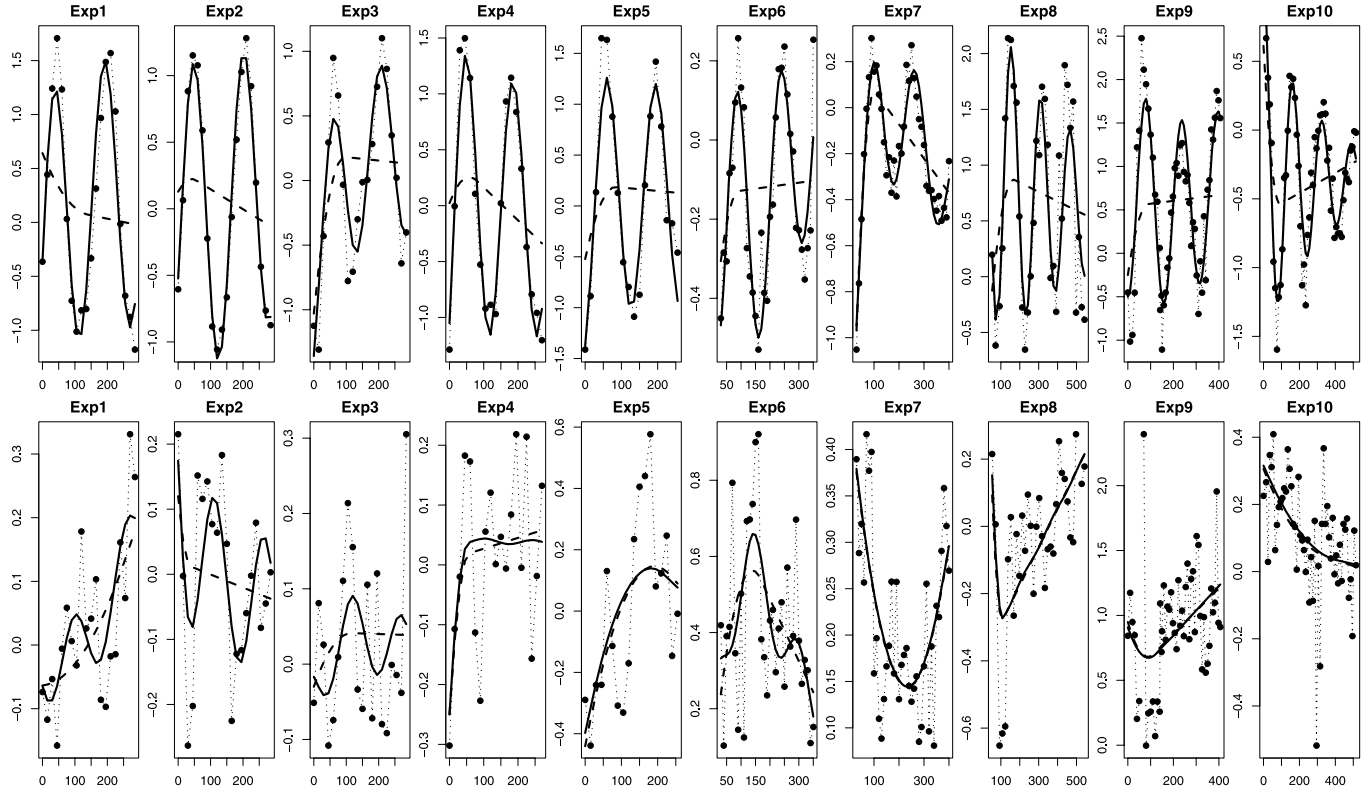}

\vspace*{-6pt}
\caption{Observed data and fitted mean curves for two samples of
genes. For each sub-figure, the horizontal axis is the time
(minutes) and the vertical axis is the gene expression value
(log-ratio). The first row of sub-figures shows the ten time
series for a known PE gene (SPAPYUG7.03C). The second row is for a
stress response gene (SPAC23C4.09C), which is not regulated by the
cell cycle. The bullet dots are the observed data. They are
connected by dotted lines. The solid lines are the mean curves
obtained by fitting the $M_{1}$ model to the data. The dashed
lines are the mean curves obtained by fitting the $M_{0}$ model to
the data. The details of model fitting are given in the following
text.} \label{SampleGene} % caption for the whole figure
\end{sidewaysfigure}

Let $Y_{get}$ denote the gene expression log-ratio at time
$T_{et}$ in experiment $e$ for gene $g$, where $g=1,\ldots, G$,
$e=1,\ldots, E$, $t=1,\ldots, S_{e}$. Here $Y_{get}$ is the
observed data; $T_{et}$, the time of the measurement; $G$, the total
number of genes studied; $E$, the total number of independent
experiments; and $S_e$, the total number of time points measured in
experiment $e$.
The whole data set can be visualized as a G-by-E matrix of time
series, where each row corresponds to one gene and each column
corresponds to one experiment. If we pool together all  filtered data
from the ten data sets, we have that $G=4994$, $E=10$, and $S_e$
ranges from 18 to 52. A detailed overview of the data is given in
Table~\ref{DataOverviewTable}. For illustration, the data for two
genes are shown in Figure~\ref{SampleGene}.

%s2.2 ###
\subsection{Model}\label{section:model}

We model each time series as a mean curve with additive
independent and identically distributed (i.i.d.) Gaussian noise
for measured time points. The mean curve is a function of time
consisting of a trend component and a periodic component. For the
trend component, we use a linear function along with a truncated
quadratic function to model the block-release effect [artifacts
introduced by experimental protocols for synchronization; see
\citet{refLuLiu04NAR}] and the general trend shown by the time
series. We assume a first order Fourier model for the periodic
component. A damping term is added to the periodic component to
model the cell cycle de-synchronization effect, which implies that
%
%f1 ###
the periodic phenomenon eventually disappears as time increases.
To model the whole matrix of time series, we assume that the
periodic components for all genes within one experiment share the
same period, which is equal to the cell division time (i.e.,
duration between the birth of a cell up to its division into two
daughter cells). We further assume that the relative peak time
within the cell cycle for every gene is fixed, which allows all
genes to share the same phase shift when the periodic components
across experiments are compared. More specifically, we assume the
following model ($M_1$) for each time series:
\begin{eqnarray*}
Y_{get}&=&a_{ge}+b_{ge}T_{et}+c_{ge}\bigl(\min(T_{et}-d_{ge},0)\bigr)^2
\\
&&{}+A_{ge}\cos(\mu_eT_{et}+\psi_e+\phi_g)e^{-\lambda_e T_{et}}+\varepsilon_{get},
\end{eqnarray*}
where
\begin{longlist}[]
\item[]\hspace*{-17pt}$a_{ge}+b_{ge}T_{et}+c_{ge}(\min(T_{et}-d_{ge},0))^2$: trend component,\vspace*{1.5pt}
\item[]\hspace*{-17pt}$A_{ge}\cos(\mu_e T_{et}+\psi_e+\phi_g)e^{-\lambda_e T_{et}}$: periodic component,
\item[]\hspace*{-17pt}$\varepsilon_{get} \sim N(0, \sigma^2_{ge})$: i.i.d. noise,\vspace*{1pt}
\item[]\hspace*{-17pt}$a_{ge}$, $b_{ge}$: coefficients of the linear trend of a time series,
\item[]\hspace*{-17pt}$d_{ge}$: ending time of block-release effect of a time series,
\item[]\hspace*{-17pt}$c_{ge}$: magnitude of block-release effect of a time series,\vspace*{1pt}
\item[]\hspace*{-17pt}$\sigma^2_{ge}$: noise level of a time series,
\item[]\hspace*{-17pt}$A_{ge}$: amplitude of periodic component of a time series,
\item[]\hspace*{-17pt}$\mu_e$: cell cycle angular frequency, equal to $2\pi$ divided by the period of cell cycle of an experiment,
\item[]\hspace*{-17pt}$\psi_e$: experiment-specific phase, which models the phase-shift between two experiments,
\item[]\hspace*{-17pt}$\phi_g$: gene-specific phase, which decides its peaking time,
\item[]\hspace*{-17pt}$\lambda_e$: magnitude of the de-synchronization effect of an experiment.
\end{longlist}

For each gene, we use different amplitude parameter $A_{ge}$ for
different experiments to account for the effects of different
experimental platforms and synchronization techniques. If a gene
is not periodic, the fitted amplitude $A_{ge}$ should be close to
zero. For such time series, the phase parameter $\phi_g$ is
redundant. To capture different noise levels in different
experiments, we specify a hierarchical structure for the noise
component by assuming that all $\sigma^2_{ge}$ from the same
experiment share the same inverse chi-square distribution with
chosen degree of freedom $C_{12}$ (a~constant specified in
the \hyperref[appendix:A]{Appendix}) and unknown hyper-parameters
$\zeta_{e}$:
\[ \sigma^2_{ge}|\zeta_{e} \sim \mathit{Inv}\mbox{-}\chi^2(C_{12}, \zeta_{e}). \]

For convenience, we introduce the following notation:
\begin{longlist}[]
\item[]\hspace*{-17pt}$Y \equiv \{Y_{get}, \mbox{ for }   g=1,\ldots, G; e=1,\ldots, E; t=1,\ldots, S_{e}\}$: expression values,
\item[]\hspace*{-17pt}$\Theta_{e} \equiv \{\mu_{e}, \psi_{e}, \lambda_{e}, \zeta_{e} \}$: experiment-specific parameters,
\item[]\hspace*{-17pt}$\Theta \equiv \{\Theta_{1},\ldots, \Theta_{E} \}$,
\item[]\hspace*{-17pt}$\Phi \equiv \{\phi_{1},\ldots, \phi_{G}\}$: gene phases,
\item[]\hspace*{-17pt}$\Gamma_{ge} \equiv \{a_{ge},b_{ge}, c_{ge}, d_{ge}, A_{ge},\sigma^2_{ge}\}$: time-series-specific parameters,
\item[]\hspace*{-17pt}$\Gamma_{g} \equiv \{\Gamma_{g1},\ldots, \Gamma_{gE}\}$,
\item[]\hspace*{-17pt}$\Gamma \equiv \{\Gamma_{1},\ldots, \Gamma_{G}\}$.
\end{longlist}

\begin{figure}

\includegraphics{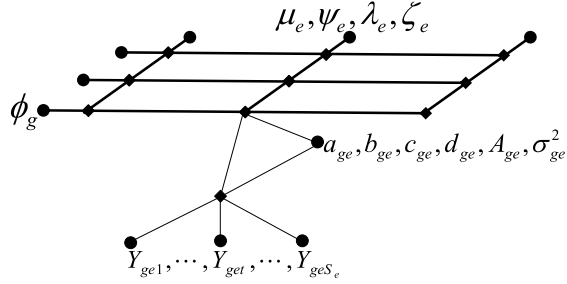}

\caption[Dependence structure of all variables]{Dependence
structure of all variables. All links are undirected. Bullets
represent a variable or a group of variables. Diamonds represent
the dependence of the variables linked to it. Corresponding to the
G-by-E matrix of time series, the main parameter structure can be
visualized as a matrix, where each row corresponds to a
gene-specific parameter $\phi_g$ and each column corresponds to
experiment-specific parameters $(\mu_e,\psi_e,\lambda_e,\zeta_e)$.
Each cell of the matrix corresponds to the variables specific to a
time series. For example, all $\phi_g$'s are independent of each
other conditional on all $(\mu_e,\psi_e,\lambda_e,\zeta_e)$; a
time series is independent of all other time series conditional on
the union of $\phi_g$ and $(\mu_e,\psi_e,\lambda_e,\zeta_e)$.}\label{ParameterStructure_v9}
\end{figure}

All variables may be visualized within a gene-by-experiment (i.e.,
$G \times E$) matrix (Figure~\ref{ParameterStructure_v9}), which
shows their dependence structure. Each row corresponds to a
gene-specific parameter $\phi_g$ and each column represents the set of
experiment-specific parameters $(\mu_e,\psi_e,\lambda_e,\zeta_e)$.
Each cell of the matrix corresponds to the variables specific to a
time series. The gene-specific parameter $\phi_g$ is the key to
integrate the time series for gene $g$ from multiple experiments.
Experiment-specific parameters $\Theta_{e}$ are used to pool
information across all genes within a particular experiment.

For model comparison, we also introduce the following model
($M_0$) for APE genes:
\[
Y_{get} = a_{ge} + b_{ge} T_{et} + c_{ge}
\bigl(\min (T_{et} -d_{ge}, 0)\bigr)^2 + \varepsilon_{get}.
\]
The only difference between $M_0$ (null model) and $M_1$
(alternative model) is the periodic component $A_{ge} \cos (\mu_e
T_{et} + \psi_e + \phi_g) e^{-\lambda_e T_{et}}$.

%f2 ###

%s2.3 ###
\subsection{Identifiability}\label{section:identifiability}

In the $M_1$ model, the phase parameters $\psi_e$ and $\phi_g$ are
not identifiable because the joint posterior distribution remains
the same if we add a constant $z$ to all $\psi_e$ and subtract $z$
from all $\phi_g$. This nonidentifiability problem can be solved
by fixing one of the phase parameters, but the loss of one degree
of freedom makes the MCMC algorithm very ``sticky'' (slow-mixing).
Since we only care about the relative values of $\psi_e$'s and
$\phi_g$'s, we solve the problem by assigning a reasonably tight
prior distribution to one of the phase parameters and flatter
priors to others, and using a transformation group move to improve
mixing of the MCMC chain (see Appendix~\ref{p2:section:Extra}).

For periodic signal fitting, the angular frequency parameter
$\mu_e$ is usually nonidentifiable because a time series with
angular frequency $\mu_e$ is also a time series with angular
frequency $\mu_e/n$ for any positive integer $n$. We avoid this
problem by specifying the periodic signal as a damping single
sinusoidal curve and limiting the domain of $\mu_e$ to a bounded
range. The bound of $\mu_e$ is instituted via its prior which is
based on our prior knowledge of the cell cycle duration in fission
yeast.

%s2.4 ###
\subsection{Bayesian computation}\label{section:computation}

We estimate all unknown parameters through MCMC simulation of
their joint posterior distribution. More specifically, we use a
Metropolis-within-Gibbs algorithm to iteratively sample one set of
parameters given all the others:
\begin{itemize}
\item Step 1: sample experiment-specific parameters $\Theta_{e}$
conditional on $\Phi$, $\Gamma$ and $Y$,

\item Step 2: sample gene-specific parameters $\phi_g$ conditional
on $\Theta$, $\Gamma$ and $Y$,

\item Step 3: sample time series-specific parameters $\Gamma_{ge}$
conditional on $\Theta$, $\Phi$ and $Y$.
\end{itemize}

The MCMC chain composed of these basic moves suffers from a slow
mixing problem caused by strong correlations among some
parameters. We can alleviate the problem by parallelizing each of
the three steps based on the conditional independence of the
parameters. For instance, we can parallelize the sampling of
$\Gamma_{ge}$ from their full conditional distribution since they
are independent of each other given $\Theta$, $\Phi$ and $Y$. When
some parameters are highly correlated in their joint distribution,
single-component moves cause very slow-mixing. To cope with this
problem, we designed a new sampler called the Metropolized
independence group sampler (MIPS) by combining the ideas of
grouping [\citet{refLiuKong94Biometrika}] and the Metropolized
independence sampler [\citet{refHastings70Biometrika}; \citet{refLiu96StatComput}; \citet{refLiu01book}]. The key idea is to update
the whole subset of correlated variables simultaneously
independent of the current state using a sequential proposing
procedure. MIPS moves are inserted to the main
Metropolis-within-Gibbs iteration. The details of the MCMC
implementation are given in the \hyperref[appendix:A]{Appendix}.

%s2.5 ###
\subsection{Strategies for discerning PE genes from APE genes}\label{section:strategy}

We used three statistics to judge which genes are PE ones. Among
them, the Bayesian Information Criterion is used to compare the
fitting of model $M_1$ with that of model $M_0$, both to real
data. The other two statistics measure the periodicity by comparing
the fitting of the $M_1$ model to the real data with that to the permuted
data or the data simulated from the $M_0$ model.

%s2.5.1 ###
\subsubsection{Permutation test}\label{section:permutation}

Since we fit model $M_1$ to every gene, even the APE genes are
modeled with experiment-specific parameters $\Theta$ that are
primarily determined by PE components. Therefore, to examine the
effect of our Bayesian model fitting procedure on APE genes, we
generate background data by permuting each time series for every
gene in the real data, which destroys any periodic pattern
therein. We run the same MCMC algorithm to fit the $M_1$ model to
the background data set by fixing all experiment-specific
parameters $\Theta$ at the posterior mode obtained from the MCMC
run for the real data.

%s2.5.2 ###
\subsubsection{Simulation from the null model}\label{section:nullmodel}

One problem of using the permutation data as background control is
that the permuted time series do not capture the intrinsic
autocorrelation of the measured time series, which exists even if
it is not periodically expressed. For example, many time series in
the real data show a general trend without oscillation, which may
be a result of the gene's response to the perturbation caused by
synchronization techniques. To accommodate this possible bias, we
generate a second data set from the $M_0$ model. Compared to the
permuted time series, $M_0$ explains the autocorrelation in the
time series by a mean curve. We run the same MCMC algorithm to fit
$M_0$ to all genes in the real data.

We simulated from the $M_0$ model a data set of similar size and
structure as the combined real data set. All parameters are
simulated from their corresponding prior distributions. Both $M_1$
and $M_0$ are fitted to this simulated data set. While fitting
$M_1$, we fix all experiment-specific parameters $\Theta$ at the
posterior mode obtained from the MCMC run for the real data.

%s2.5.3 ###
\subsubsection{Model comparison}\label{section:modelcomparison}

One approach for discerning PE genes from APE genes is to use
permuted data or data simulated from the null model as background
control, and to fit the $M_1$ model to both the real data and the
background data. The fitting of the background data is then used
to determine a threshold for the desired false positive rate
(FPR). Another approach is to fit both models $M_1$ and $M_0$ to
the real data, and then do the classification based on a
comparison of the fitness of the models. Various information
criteria can be used for this  task, such as Akaike's Information
Criterion (AIC) [\citet{refAkaike73ICSS}], the Bayesian Information
Criterion (BIC) [\citet{refSchwarz78AnnaStat}] and the Deviance
Information Criterion (DIC)
[\citet{refSpiegelhalterVanDerLinde02JRSSB}], to just name a few.

A full Bayesian alternative to our approach here is to introduce a
latent variable $I_{g}$ for each gene to indicate whether it comes
from $M_1$ or $M_0$. Then, the reversible-jump strategy
[\citet{refGreen95Biometrika}] can be used to build a MCMC
sampler to traverse the joint space of the latent indicators and
model parameters. But due to the global nature of many parameters
in our model, this approach is computationally extremely
expensive. Additionally, the results so obtained may be too
sensitive to our model assumptions. Thus, we feel that using
randomization and null model approaches in the spirit of posterior
predictive model checking [\citet{refGelmanStern96StatSinica}]
provides a more robust detection of PE genes.

%s2.5.4 ###
\subsubsection{Statistics for periodicity}\label{section:statistics}

We use multiple gene-specific statistics to measure the
periodicity of a gene. Based on the fitted parameter values for
the $M_1$ model, we define the gene-specific Signal-to-Noise Ratio
($\mathit{SNR}$) as the relative strength of the fitted periodic component
compared to the noise level:
\[
\mathit{SNR}_{g} = \sum^{E}_{e=1} \frac{
\sum^{S_e}_{t=1}\{A_{ge} \cos (\mu_e T_{et} + \psi_e + \phi_g)
e^{-\lambda_e T_{et}} \}^2 }{\sigma^2_{ge}}.
\]
The $\mathit{SNR}$
statistic combines periodicity information for a gene from every
experiment in terms of the amplitude of its periodic component.
For each gene, we calculate $\mathit{SNR}$ for each iteration of the MCMC
chain, and then summarize the posterior samples of $\mathit{SNR}$ using the
2.5th percentile, the 97.5th percentile and the mean. Genes with
higher $\mathit{SNR}$ values are more likely to be periodically expressed.
We also use the fitted phase to measure periodicity from the
fitted parameters of the $M_1$ model. More specifically, we use
the length of the 95\% central posterior interval (denoted by LPI)
of a gene's relative phase $\phi_g+\psi_1$ ($\psi_1$ is chosen
arbitrarily since only the difference of relative phase matters)
as one of the periodicity measures. Genes with higher LPIs are
less likely to be periodic either because their periodic
components are too weak or their multiple time series might show
inconsistent peaking time within the cell cycle.

We use the Bayesian Information Criterion difference ($\mathit{BIC}^{01}$) to
measure periodicity based on the fitted posterior modes of the two
models. Let $L^{0}_g$ and $L^{1}_g$ denote the likelihood values
for gene $g$ at the posterior mode of the parameters for models
$M_0$ and $M_1$, respectively. The model comparison criterion
$\mathit{BIC}^{01}$ is defined as
$\mathit{BIC}^{01}_g = 2\log(L^{1}_g) - 2\log(L^{0}_g) - (k_1-k_0) \log(N)$,
where $N$ is the number of
observed data points for the gene, and $k_0$ and $k_1$ are the number
of free parameters in models $M_0$ and $M_1$, respectively. A gene\vspace*{1pt}
with positive $\mathit{BIC}^{01}$ value prefers model $M_1$ to $M_0$. Genes
with higher $\mathit{BIC}^{01}$ values are more likely to be periodically
expressed.

%s3 ###
\section{Results and discussion}

%s3.1 ###
\subsection{Model fitting check}

The MCMC chain on the entire real cell cycle data converged in
approximately 2000 iterations. The autocorrelation function of the
posterior probabilities from each chain showed that the MCMC
algorithm is efficient in terms of effective sample sizes after
burn-in. The details of the model fitting diagnosis are given in
the supplemental material of this paper
[\citet{refFanLiu09AOASsup}]. Figure~7 in the supplementary
material [\citet{refFanLiu09AOASsup}] displays the posterior
distribution of the cell cycle length $2\pi/\mu_e$ for each of the
ten experiments. After convergence, the experiment-specific
parameters $\Theta$ showed little variation, that is, their marginal
posterior distributions had very small variance compared to their
ranges. Based on the posterior mode determined from the MCMC
chain, we calculated the residue of each time series. The
autocorrelation analysis of the residue showed that by fitting
$M_1$ to the data, the autocorrelation was reduced to the level
comparable to those of i.i.d. noise. Comparison of variance
reduction between the real and the permuted data suggested that
the $M_1$ model explained a significant amount of variance for
most of the genes showing significant autocorrelation in their
time series.

%f3 ###
\begin{figure}

\includegraphics{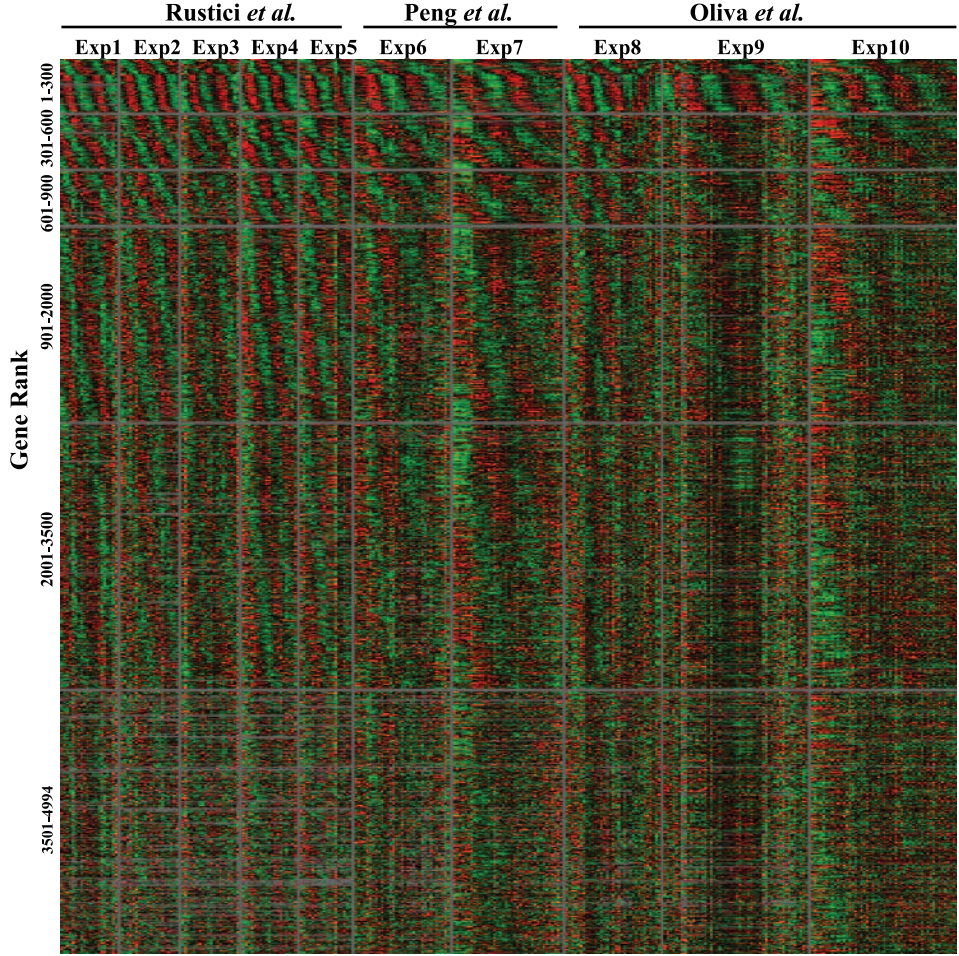}

\caption{Heatmap of all genes' time series data
ranked by decreasing mean SNR value. Columns correspond to time
points, which are grouped by experiment and sorted by time within
each group. Rows correspond to genes, which are ranked by their
mean SNR value and sorted by their mean peak times within each
group. For example, the first row group contains the 300 genes
with the highest mean SNR value from our combined analysis of all
10 experiments, and they are sorted by their relative phase
$\phi_g+\psi_1$ within the group. Each time series is normalized
to zero mean and unit variance for display. The heatmap is drawn
by TreeView [Eisen et~al. \protect(\citeyear{refEisenBotstein98PNAS})] with default
setting. Red indicates up-regulation, green indicates
down-regulation, black means no change of expression levels, and
grey is missing data. It shows a periodic pattern for all gene
groups.} \label{All_Heatmap}
\end{figure}

%s3.2 ###
\subsection{Number of periodically expressed genes}

We ranked all genes in the order of decreasing posterior mean
$\mathit{SNR}$ value. Thus, highly ranked genes are more likely to be
periodically expressed. We then stratified this sorted list into 6
groups and reordered each group according to the fitted peaking
time. Figure~\ref{All_Heatmap} shows the whole sorted data set.
Strikingly, a periodic pattern stands out for all gene groups
after simply reordering them (note that these are simply
rearranged original data). The pattern is clear and consistent
across all experiments for the top 2000 genes, which suggests that
about 40\% of all genes in the organism could be periodically
expressed. The pattern is still strong for genes in the range
2001--3500. We can even observe periodicity among the remaining
genes shown in the bottom group, which, however, is comparable to
the top ranking ``genes'' in the permuted data.

For a comparison with the result from traditional clustering
methods, the microarray clustering software Cluster
[\citet{refEisenBotstein98PNAS}] was used to group genes with
similar gene expression. A heatmap similar to
Figure~\ref{All_Heatmap} is included in the supplemental material of
this paper [\citet{refFanLiu09AOASsup}]. Compared to the
ubiquitous periodic pattern in Figure~\ref{All_Heatmap}, only several
small clusters with visible periodic pattern may be observed from
the hierarchical clustering result.

%f4 ###
\begin{figure}[b]

\includegraphics{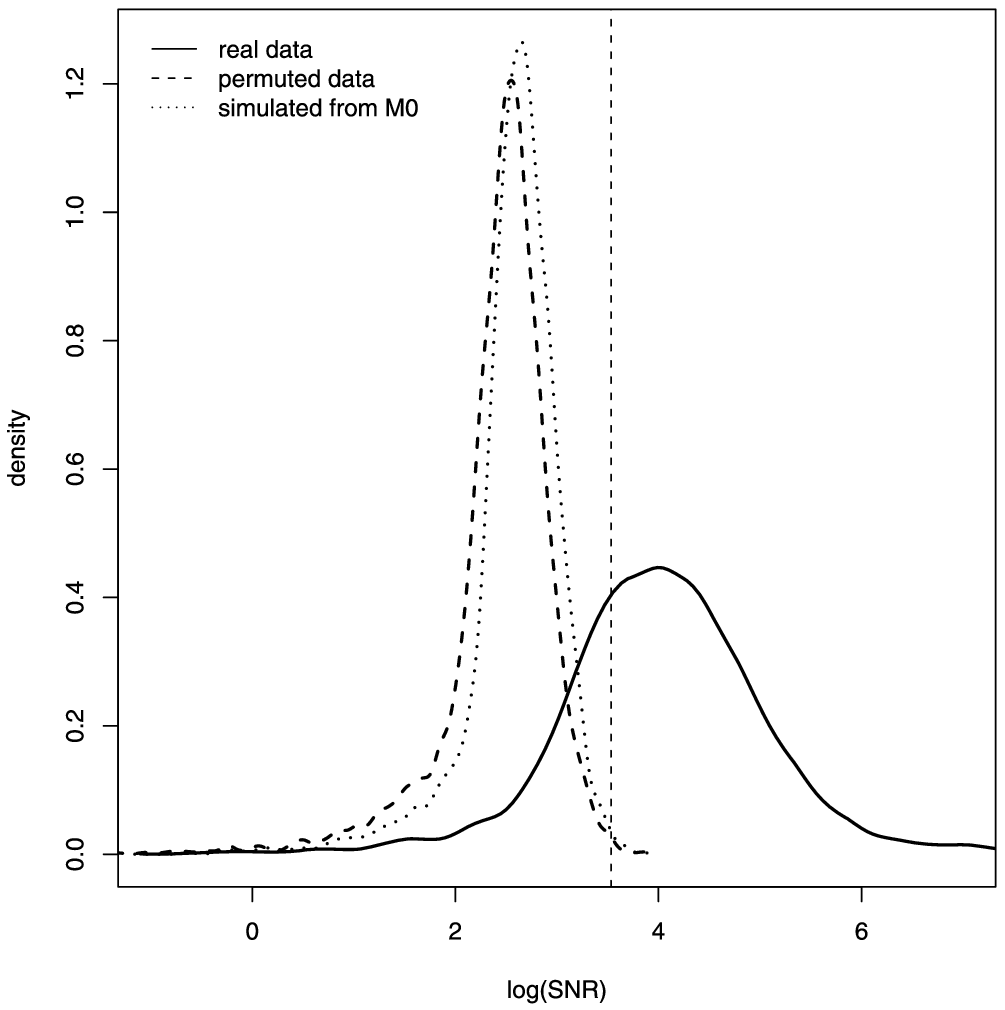}

\caption{Density comparison of SNR from the three data sets. The
$M_1$ model is fitted to the real data, permuted data and the
data simulated from the $M_0$ model. For each gene, we get the
posterior mean of the SNR statistic from the combined analysis.
For each data set, we pool all genes together to get a kernel
density estimate, which is shown in this graph. The vertical line
indicates the threshold corresponding to FPR${}={}$0.002 in the permuted
data, from which one can claim 3599 PE genes from the real data.}\label{Fig_SNRDensityComparison}
\end{figure}

%t2 ###
\begin{table}
 \caption{Correlation of different statistics and their classification results}\label{StatisticsComparisonTable}
\begin{tabular*}{8.2cm}{@{\extracolsep{\fill}}lccc@{}}
\hline
\textbf{Statistic} & \textbf{SNR} & \textbf{LPI} & $\bolds{\mathit{BIC}^{01}}$  \\
\hline
SNR                             & 3599 & 3051 & 1967  \\
LPI                             & $-$0.93 & 3086 & 1906  \\
$\mathit{BIC}^{01}$     & \phantom{$-$}0.86 & $-$0.83 & 2003  \\
\hline
\end{tabular*}
\tabnotetext[]{tm}{\textit{Note}: The permuted data was used as background control. The
lower-left part of the table shows the Spearman correlation
between pairs of statistics. The numbers on the diagonal are the
number of PE genes claimed by the corresponding statistic. For SNR, we
use a cutoff corresponding to FPR${}={}$0.002 for the two mean SNR
density. For LPI, we also use the threshold corresponding to
FPR${}={}$0.002. We use zero as the threshold for $\mathit{BIC}^{01}$. The
upper-right part of the table shows the number of PE genes claimed
by a pair of statistics. Within them, 1898 genes are claimed by
all three statistics.}
\end{table}

We used two approaches to test whether the visual periodic pattern
in Figure~\ref{All_Heatmap} is statistically significant. The first
approach compares the fitting of  the $M_1$ model to the real and
background data, that is, the permuted data or the data simulated
from the $M_0$ model. Two statistics are used to measure the
periodicity for this approach. The SNR statistic measures the
amplitude of the periodic component, while the LPI statistic
measures the uncertainty of the relative phase of every gene.
Figure~\ref{Fig_SNRDensityComparison} and Figure~8
in the supplementary material [\citet{refFanLiu09AOASsup}] show the estimated
posterior densities of these measures. The curves from the
background data provide a null distribution for the corresponding
statistic, from which we can estimate FPR for any given threshold.
The clear separation of the posterior densities for the real and
background data suggests that a lot of genes show a periodic
pattern that is stronger than i.i.d. noise or $M_0$ data. For
example, by comparing the LPI curves of the real and permuted data
in Figure~8 in the supplementary material
[\citet{refFanLiu09AOASsup}], we can claim 3086 PE genes for
FPR${}={}$0.002, corresponding to about 10~false positives. Similarly,
by comparing the posterior mean SNR values of the real and
permuted data in Figure~\ref{Fig_SNRDensityComparison}, we can claim
3599 PE genes for FPR${}={}$0.002. The number of claimed PE genes when
using the simulated data from the $M_0$ model as background
control is similar. For instance, the comparison of the posterior
mean SNR densities yields 3414 PE genes for FPR${}={}$0.002, and that of
the LPI densities yields 3036 PE genes for FPR${}={}$0.002.

The second approach compares the fitting of the two models $M_1$
and $M_0$, both using the real data. We used BIC as the model
comparison criterion. As shown in Figure~9 in the supplementary
material [\citet{refFanLiu09AOASsup}], almost all $\mathit{BIC}^{01}$
values from the permuted data as well as the simulated data from
the $M_0$ model are smaller than zero. For the real data, we can
claim 2003 PE genes from the combined analysis by using zero as
the threshold for $\mathit{BIC}^{01}$. Corresponding to this threshold, the
permuted data will only produce one false positive PE gene,
corresponding to FPR${}={}$0.0002.

The results of these three statistics are summarized in
Table~\ref{StatisticsComparisonTable}. Here we used the permuted
data as background control. The average Spearman correlation
between pairs of the statistics is 0.87, suggesting that the
three statistics are highly consistent in ranking the genes' periodicity.
The approaches based on permutation control (SNR, LPI) made more
significant claims than the model selection approach. Overall, we
obtained a list of 1898 significant PE genes that are claimed by all three statistics.
% the
%three statistics agree on the significantly periodically expressed
%genes leading to claimed by these three statistics are highly
%consistent, producing a list of 1898 genes that are claimed by
%all.

%s3.3 ###
\subsection{Performance comparison}

To evaluate the performance of identifying PE genes, we defined a
benchmark set as the union set of the list of PE genes derived
from small-scale experiments [\citet{refMargueratBahler06Yeast}]
and a core set of genes whose periodic regulation is conserved
between budding yeast and fission yeast
[\citet{refLuBarJoseph07GenomeBiol}]. The resulting benchmark
set consists of 162 genes. We used this benchmark set to compare
our method with the method used by
\citet{refMargueratBahler06Yeast}.

%f5 ###
\begin{figure}

\includegraphics{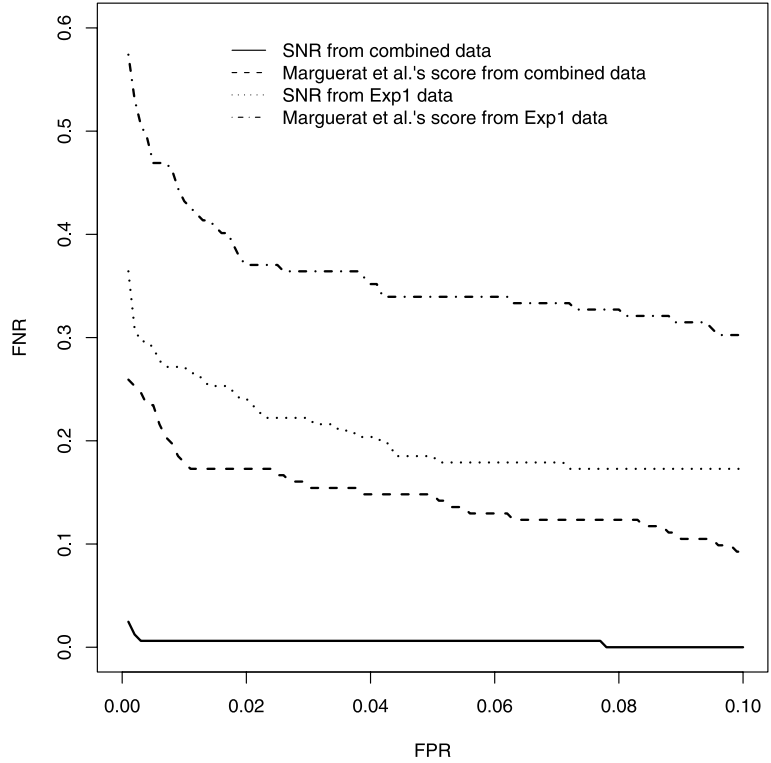}

\caption{Performance on the
benchmark set. For each of the four methods listed in the figure
legend, we plot FNR against FPR under various thresholds. For each
threshold, the benchmark set of 162 PE genes is used to estimate
FNR. The permuted version of the data is used to estimate FPR. A
smaller under-curve area corresponds to a better classification
performance for the benchmark set.}\label{ROCComparisonWithMarguerat}
\end{figure}

The statistic used for gene classification by
\citet{refMargueratBahler06Yeast} is a score calculated from a
$p$-value of regulation and a $p$-value of periodicity. When combining
multiple experiments for gene classification, they multiplied the
$p$-values from individual experiments to get a total $p$-value of
regulation and a total $p$-value of periodicity. To estimate the FPR
of their statistic, we calculated the scores for the permuted
data. For our method, we use the SNR statistic for gene
classification.

Figure~\ref{ROCComparisonWithMarguerat} shows the performance of the SNR statistic and Marguerat et al.'s score on both
the combined data (all experiments) and the Exp1 data (a single
experiment) in the form of ROC curves. For any given FPR value, we
estimate the threshold of a statistic from the permuted version of
the data. The corresponding false negative rate (FNR) is estimated
by the fraction of the genes in the benchmark set that are
classified as APE gene according to this threshold. When applied
on the data from a single experiment (Exp1 data), the SNR
statistic apparently outperforms Marguerat et al.'s
score. The gain of statistical power at the single experiment
level could be due to our explicit modeling of the trend component
and the de-synchronization effect, which makes our model more
realistic for the cell cycle time series. When comparing their
performances on the combined data, it seems that the SNR statistic
increases the statistical power over Marguerat et al.'s
score significantly. This is due not only to a more realistic
model for single time series, but also to our approach of the
Bayesian meta-analysis. Instead of combining the $p$-values from
individual experiments, we model multiple experiments
simultaneously so as to borrow information across experiments.

Figure~\ref{ROCComparisonWithMarguerat} indicates that the same
statistic performed better at discerning PE genes with the
combined data than with the data from a single experiment. This is
also true when comparing the performances of a statistic using the
overall combined versus that using any subset of the experiments.
The detailed information is given in Table~2 in the supplementary
material [\citet{refFanLiu09AOASsup}]. This is natural because
any subset contains less information than the full combined data;
but on the other hand, it also indicates that each experiment
captured some information about genes' periodicity during the cell
cycle.

%s3.4 ###
\subsection{Subset analysis}

\begin{figure}

\includegraphics{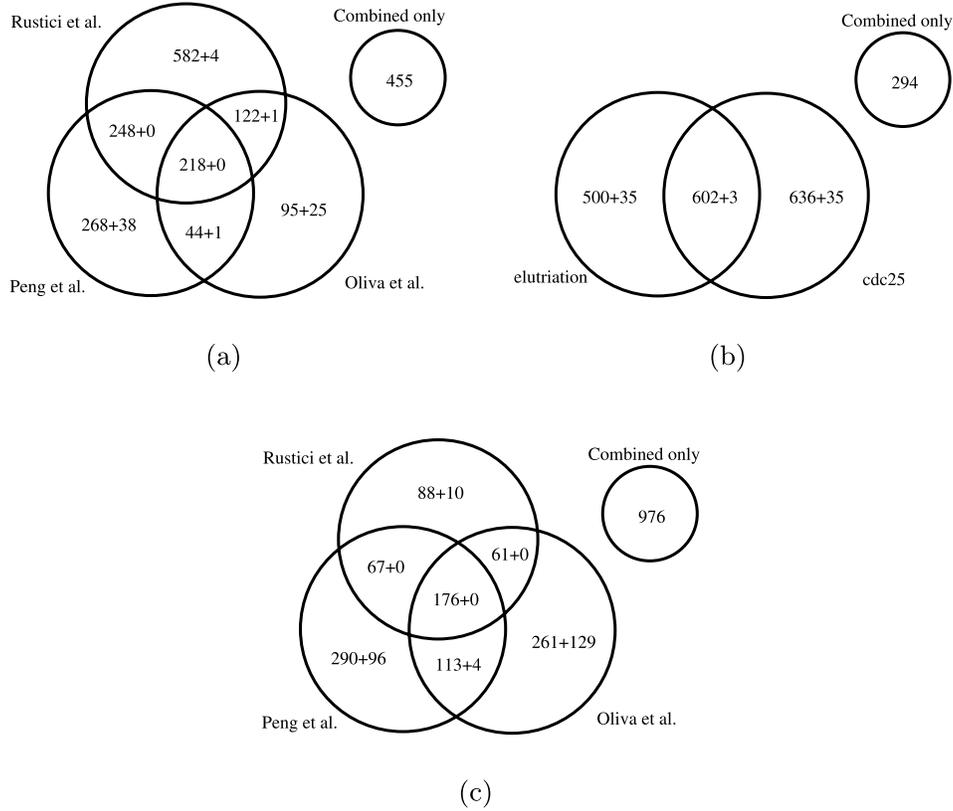}

\caption{Venn diagrams showing overlap between
claimed PE genes from subsets of the data. Each gene set in all
diagrams is compared with the result from the combined analysis
that we did using our method. The number before the plus sign is
the number of genes also claimed as periodically expressed by our
combined analysis. The stand-alone circle represents the part
which is reported only by the combined analysis. \textup{(a)} Comparing the
results from individual data sets using our method. \textup{(b)} Comparing
the results from two synchronization techniques using our method.
\textup{(c)} Comparing the results reported in original studies.}\label{VennDiagram} % caption for the whole figure
\end{figure}

To compare three individual studies
[\citet{refRusticiBahler04NatGenet}; \citet{refOlivaLeatherwood05PLoSBiol}; \citet{refPengLiu05MolBiolCell}]
and different experimental techniques, we used the same method for
the combined data set to fit model $M_1$ to all three individual
data sets, and also the two collections of experiments using different
synchronization techniques (elutriation or cdc25 block-release).
We first determined the 95\% posterior interval of the SNR
statistic for each gene to account for the uncertainty of its SNR
estimate. Then for comparison of all the subsets at the same
significance level, we claim a gene to be PE if its posterior mean
SNR value is above the upper 97.5\% posterior limits of the SNR of
at least 4984 (out of 4994) permuted ``genes.'' For the combined
data, we thus claimed 2032 PE genes. Figure~\ref{VennDiagram}(a) and
Figure~\ref{VennDiagram}(b) show the overlap of the results from our
subset analyses. Figure~\ref{VennDiagram}(c) shows the overlap of the
original results from the three individual studies. There are 976
genes which are reported as PE by our combined analysis but not by
any of the three original studies. Supporting evidences for these
genes are included in the supplementary material
[\citet{refFanLiu09AOASsup}].

%f6 ###

Similar to Figure~\ref{VennDiagram}(c), the discrepancy about the count
and identity of PE genes exists between individual data sets
[Figure~\ref{VennDiagram}(a)] and across synchronization techniques
[Figure~\ref{VennDiagram}(b)] although we have unified the whole
analysis procedure. Therefore, instead of attributing the
discrepancy between the subsets to inconsistent gene naming or use
of different analysis methods or arbitrary thresholds
[\citet{refMargueratBahler06Yeast}], we suggest that the cause
is intrinsic to the data. It also shows that most genes in the
discrepant part show significant periodicity in the combined
analysis. The combined analysis also captured many genes which can
not be detected by subset data analysis. Combined with the
benchmark analysis, we observed that 5 out of the 40 benchmark
genes whose periodicity have been confirmed by small-scale
experiments [\citet{refMargueratBahler06Yeast}] were missed by
all three original studies as well as our combined analysis. On
the other hand, 6 out of the 92 core environmental stress response
genes with known function [\citet{refChenBahler03MolBiolCell}]
were claimed as periodically expressed by all three original
studies as well as by our combined analysis, suggesting that their
periodic signal is clear to all methods. Possibly, the periodicity
measure for widely used positive or negative benchmark sets are
not quite accurate.

To investigate the discrepancy between different subsets, we
systematically tested these subsets' pairwise reproducibility
using the posterior mean SNR values. If it is true that the genes
have an intrinsic order in terms of periodicity and all individual
data sets are of similar quality in revealing this ordering
information, the periodicity measures across pairs of subsets
should be consistent. Each data set yields a SNR vector measuring
the periodicity of all genes. The key idea is to check whether the
Spearman correlation of the two SNR vectors is still significant
after removing genes which are top ranked in both vectors. The
details are shown in Figure~10 in the supplementary material
[\citet{refFanLiu09AOASsup}]. After removing the 847 genes that
are highly ranked by both Peng et al. and Oliva
et al., the remaining genes' SNR values from these two
data sets show no positive Spearman correlation at the
significance level of 0.05. This sets the number of reproducible
genes supported by these two data sets (5 experiments) to 847.
This same count increases to 934 for Rustici et al.
versus Peng et al. (7 experiments), and to 1008 for
Rustici et al. versus Oliva et al. (8
experiments). The increasing of reproducible genes is consistent
with the increase in the size of data involved in comparison. The
number further increases to 1554 when comparing elutriation
experiments with cdc25 experiments. This suggests that although
the number of reproducible genes is less than the number of PE
genes suggested by the combined analysis, the reproducibility is
improved by including more data in the comparison or by
partitioning the data according to the experiment technique.

To explain the above subset discrepancy, possible flaws in the
benchmark sets and the high number of significant genes in the
combined analysis, we hypothesize a network-based dynamics for the
cell cycle process. For instance, periodic signals from
transcription of key cell cycle-regulated genes propagate through
the relevant downstream regulatory networks of the organism
potentially targeting a considerable number of genes. Thus,
depending on the status of the network, these genes may show an
observable periodic pattern under one condition, and be too weak
to detect under another condition. As a consequence of the
combined effect of the variation in periodicity and experimental
noise, each study could capture a different subset of the PE
genes. The difference of the cell cycle length shown in Figure~7 in
the supplementary material [\citet{refFanLiu09AOASsup}], which
could not be explained solely by microarray platform difference,
is a further evidence of such variation. For example, the cell
cycle lengths in the posterior mode for the two cdc25 experiments
in Rustici et al. are 135 and 138 minutes, while in Oliva
et al. and Peng et al., this number increases to
164 and 173 minutes, respectively. Although they are using the
same synchronization technique on the same organism, subtle
environmental or physiological differences have changed the speed
of the cell cycle oscillation. Therefore, it may have also changed
relative amplitudes of oscillation of the genes leading to overall
ranking discrepancy.

%s4 ###
\section{Conclusion}

In spite of the rapid rise in the number of microarray
experiments, many of which address related issues, a systematic
meta-analysis of such data is rarely attempted. We conducted a
meta-analysis of ten fission yeast cell cycle genome-wide
time-series experiments with a model-based Bayesian approach.
Compared to other methods, key features of our model include the
fixed relative phase of the peaking time of the genes across all
experiments (e.g., a gene will peak 10 degrees earlier than
another gene in an experiment if and only if the same happens in
another experiment) and a flexible amplitude for periodic
components. Our approach does not require training sets to
estimate important global parameters such as the period of cell
cycle, but to infer them from all the data. Notably, our parametric
approach deals with phase shift,
signal amplitude difference, noise level difference and
de-synchronization automatically. Despite the high
dimensionality, the implemented MCMC chain mixes well with the
help of global moves. The residual analysis shows that our model
fits the data well.

A striking finding of our analysis is that more than 2000 genes
are significantly periodically expressed, which accounts for
approximately 40\% of all the genes in the fission yeast genome.
The subset analysis suggests that this number may increase with
more data included. This enhances greatly the current knowledge of
only 10--15\% of all fission yeast genes that are reported as
periodically expressed during the cell cycle. Interestingly,
genome-wide oscillation has also been reported by recent studies
on other cyclic phenomena in the cell, such as the metabolic cycle
and circadian periodicity [\citet{refKleveczMurray04PNAS};
\citet{refTuMcKnight05Science}; \citet{refPtitsynGimble07PLoSCompBiol}].
Clearly, a certain amount of influence of the global cell cycle
processes on most genes in the genome, in particular, in
unicellular organisms such as fission yeast, cannot be ruled out.
For instance, the folding and unfolding of chromosomes over the
course of the cell cycle will have genome-wide incidental effect on
transcription. However, earlier studies concede that limited
ability to distinguish precisely the weakly periodic oscillations
from prevalent microarray noise only allowed conservative
estimates of PE genes. By explicitly modeling periodic and
nonperiodic components, and different sources of variation and
noise, our model-based approach helps to overcome this
long-standing limitation. The resulting list of more than 2000 PE
genes would allow the researchers to cast a much wider and deeper
net for cell cycle regulated genes that can lead to investigation
of novel or relatively less known gene modules and networks
involved in the machinery of cell cycle regulation.

It should be noted that the key idea behind our model is rather
general. It can be applied to detect periodic patterns where the
amplitude is noisy but the patterns are nonetheless consistent
across different experiments. The data can be any collection of
time series. A study of cell cycle data from other species, such as
the budding yeast, mouse, human, etc., using the proposed method
can be of immediate interest.

One possible way to improve the current method is to employ a more
robust error model, using, for example, $t$-distributions instead of
Gaussians for the noise term [\citet{refHampelStahel86book}; \citet{refLangeTaylor89JASA}]. But as a price to pay, the
computational complexity may be increased substantially. It should
be noted that, as stated in Section~\ref{section:modelcomparison}, alternative Bayesian
model selection methods may also be applied to this problem. For
example, \citet{refGreen95Biometrika} provides a way to perform
joint model selection and parameter estimation via reversible jump
MCMC. It may be applicable to this problem if the efficiency of
reversible jump MCMC moves can be improved significantly. The
methods proposed by \citet{refChib95JASA} and
\citet{refChibJeliazkov01JASA}, which estimate the marginal
likelihood of the data under a model, may also be a worthwhile
direction to explore.

\begin{appendix}
%s5 ###
\section*{Appendix: MCMC implementation}\label{appendix:A}

%s5.1 ###
\subsection{Prior distribution}\label{p2:section:priors}

We assigned reasonably diffuse but still proper prior
distributions for all parameters:
\begin{eqnarray*}
a_{ge} &\sim&  N(0, C_{1}), \\
b_{ge} &\sim&  N(0, C_{2}), \\
c_{ge} &\sim&  N(0, C_{3}), \\
d_{ge} &\sim&  \operatorname{Unif}(0, C_{4}), \\
A_{ge} &\propto&  \operatorname{Exp}(\mathit{rate}=C_{5}),        \qquad  0 \leq A_{ge} < C_{6}, \\
\mu_{e} &\sim&  \operatorname{Unif}(C_{7}, C_{8}), \\
\psi_{1} &\propto&  N(0,C_{9}),                                         \qquad  -\pi \leq \psi_{1} < \pi, \\
\psi_{e} &\propto& N(0,C_{10}),                                         \qquad e=2,\ldots, E, -\pi \leq \psi_{e} < \pi, \\
\phi_{g} &\sim&  \operatorname{Unif}(-\pi, \pi), \\
\lambda_{e} &\sim&  \operatorname{Unif}(0, C_{11}), \\
\sigma^2_{ge}|\zeta_{e} &\sim&  \mathit{Inv}\mbox{-}\chi^2(C_{12}, \zeta_{e}),\\
\zeta_{e} &\sim&  \operatorname{Exp}(C_{13}).
\end{eqnarray*}
The constants in  the prior distributions are assigned
correspondingly, making use of our prior knowledge:
$C_{1} = 1,  C_{2} = 0.005^2, C_{3} = 0.0001^2, C_{4} = 500,\break
C_{5} = 10, C_{6} = 10, C_{7} = 2 \pi /180, C_{8} = 2 \pi /120,
C_{9} = 0.2^2, C_{10} = 1^2, C_{11} = 0.006,\break C_{12} = 4, C_{13} =
50. $

%s5.2 ###
\subsection{Posterior distributions and Metropolis-within-Gibbs}\label{p2:section:PostDistr}

 We can write
down the joint distribution of the data and parameters as
\begin{eqnarray*}
&&p(Y, \Theta, \Phi, \Gamma)
\\
&&\qquad= p(Y|\Theta,\Phi,\Gamma)p(\Theta,\Phi,\Gamma) \\
&&\qquad= \Biggl[ \prod^{G}_{g=1} \Biggl\{ \prod^{E}_{e=1}
\Bigg\langle \prod^{S_e}_{t=1}
p(Y_{get}|a_{ge},b_{ge},c_{ge},d_{ge},A_{ge},\sigma^2_{ge},\phi_{g},\mu_{e},\psi_{e},\lambda_{e})\Bigg\rangle \\
&&\qquad\quad\hspace*{83pt}{}\times p(a_{ge})p(b_{ge})p(c_{ge})p(d_{ge})p(A_{ge})p(\sigma^2_{ge}|\zeta_{e})\Biggr\}
\\
&&\qquad\quad\hspace*{255pt}{}\times p(\phi_{g}) \Biggr] \\
&&\qquad\quad{}\times \Bigg\langle \prod^{E}_{e=1}p(\mu_{e})p(\psi_{e})p(\lambda_{e})p(\zeta_{e})\Bigg \rangle.
\end{eqnarray*}

We assume that all missing data are missing completely at random,
so their corresponding components are simply omitted from this
expression.

Again, we introduce the following symbols for convenience:
\begin{eqnarray*}
D_{get} &\equiv& Y_{get} - a_{ge} - b_{ge} T_{et} - c_{ge}\bigl(\min (T_{et} - d_{ge},0)\bigr)^2,
\\
 R_{get} &\equiv&  D_{get} - A_{ge} \cos (\mu_e T_{et} + \psi_e +\phi_g) e^{-\lambda_e T_{et}},
\\
X_{get} &\equiv& \bigl(1,   T_{et},  [\min(T_{et}-d_{ge},0)]\bigr),
\\
X_{ge} &\equiv&
\pmatrix{
X_{ge1} \cr
\vdots \cr
X_{geS_{e}}
},
\\
Z_{get} &\equiv& Y_{get} - A_{ge} \cos (\mu_e T_{et} + \psi_e +
\phi_g) e^{-\lambda_e T_{et}},
\\
Z_{ge} &\equiv&
\pmatrix{
Z_{ge1} \cr
\vdots \cr
Z_{geS_{e}}
},
\\
V &\equiv&
\left[
\matrix{
\frac{1}{C_{1}} &  &  \cr
  & \frac{1}{C_{2}} & \cr
  &   & \frac{1}{C_{3}}
}  \right] .
\end{eqnarray*}
From the joint distribution, we can get all full
conditional posterior distributions:
\begin{eqnarray*}
\pmatrix{
a_{ge} \cr
b_{ge} \cr
c_{ge}
} \Big| \mathit{rest}
&\sim& N \biggl(\biggl({\frac{X^{T}_{ge}X_{ge}}{\sigma^2_{ge}}}+ V\biggr)^{-1}
\frac{X^{T}_{ge}Z_{ge}}{\sigma^2_{ge}},
\biggl({\frac{X^{T}_{ge}X_{ge}}{\sigma^2_{ge}}}+ V\biggr)^{-1}\biggr),
\\
p(d_{ge}|\mathit{rest}) &\propto& \frac{1}{C_{4}} \exp
\biggl\{ -\frac{\sum^{S_{e}}_{t=1} R^2_{get} }{2 \sigma^2_{ge}}\biggr\},
\\
A_{ge}|\mathit{rest} &\propto& N(\mu,\sigma^2), \qquad  0 \leq A_{ge}< C_6 ,
\end{eqnarray*}
where
\begin{eqnarray}
\mu &=&  {\frac{ \sum^{S_{e}}_{t=1}
\cos(\mu_e T_{et} + \psi_e + \phi_g) e^{-\lambda_e T_{et}}
D_{get}-\sigma^2_{ge} C_{5}}{\sum^{S_{e}}_{t=1} \{\cos(\mu_e
T_{et} + \psi_e + \phi_g) e^{-\lambda_e
T_{et}}\}^2}},\nonumber
 \\
\sigma^2 &=&
 {\frac{\sigma^2_{ge}}{\sum^{S_{e}}_{t=1} \{\cos(\mu_e
T_{et} + \psi_e + \phi_g) e^{-\lambda_e T_{et}}\}^2}},\nonumber
\\
 p(\mu_e|\mathit{rest}) &\propto& \frac{1}{C_8 - C_7} \prod^G_{g=1} \prod^{S_e}_{t=1} \exp \biggl\{ -\frac{R^2_{get}}{2\sigma^2_{ge}}
\biggr\},\qquad  C_7 \leq \mu_e < C_8,\nonumber
\\
 p(\psi_e|\mathit{rest}) &\propto& C^{-0.5}_{9} \prod^G_{g=1} \prod^{S_e}_{t=1} \exp \biggl\{
 -\frac{R^2_{get}}{2\sigma^2_{ge}} - \frac{\psi^2_e}{2C_{9}}
\biggr\}, \qquad -\pi \leq \psi_e < \pi,   \mbox{ for }   e=1,\nonumber
\\
 p(\psi_e|\mathit{rest}) &\propto& C^{-0.5}_{10} \prod^G_{g=1} \prod^{S_e}_{t=1} \exp \biggl\{
 -\frac{R^2_{get}}{2\sigma^2_{ge}} - \frac{\psi^2_e}{2C_{10}}
\biggr\},\nonumber
\\
\eqntext{ -\pi \leq \psi_e < \pi,   \mbox{ for }   e=2,\ldots,E,\nonumber}
\\
 p(\phi_g|\mathit{rest}) &\propto& \prod^G_{g=1} \prod^{S_e}_{t=1} \exp \biggl\{
 -\frac{R^2_{get}}{2\sigma^2_{ge}}
\biggr\},\qquad  -\pi \leq \phi_g < \pi,\nonumber
\\
 p(\lambda_e|\mathit{rest}) &\propto& \prod^G_{g=1} \prod^{S_e}_{t=1} \exp \biggl\{
 -\frac{R^2_{get}}{2\sigma^2_{ge}} \biggr\},
 \qquad 0 \leq \lambda_e < C_{11},\nonumber
\\
\sigma^2_{ge} &\sim& \mathit{Inv}\mbox{-}\chi^2\biggl(S_{e}+C_{12},   \frac{C_{12}
\zeta_e + \sum^{S_{e}}_{t=1} R^2_{get}  }{S_{e}+C_{12}}\biggr),\nonumber
\\
\zeta_e &\sim& \operatorname{Gamma}\Biggl(\frac{C_{12}}{2}G+1,   \frac{C_{12}}{2}
\sum^G_{g=1} \frac{1}{\sigma^2_{ge}}+C_{13}\Biggr).\nonumber
\end{eqnarray}

For conditional distributions which we only know up to a
normalization constant, we used the Metropolis--Hastings algorithm to
draw samples. When fitting the $M_0$ model to a gene, the full
conditional distribution of its parameters can be obtained by
simply replacing all $A_{ge}$ with zero in the corresponding full
conditional distribution from $M_1$.

%s5.3 ###
\subsection{Advanced MCMC moves for better mixing}\label{p2:section:Extra}

Besides the basic Metro\-polis-within-Gibbs iteration, we insert the
following moves to perturb the MCMC chain in order to help it
traverse faster through the high dimensional space where there are
many local modes and strong correlations among a group
of parameters.
\begin{itemize}
 \item Phase parameters $\psi_e$ and
$\phi_g$ are not identifiable in model $M_1$ because the joint
posterior distribution is invariant if we add a value to all
$\psi_e$ and subtract the same value from all $\phi_g$. One way to
solve this nonidentifiability problem is to fix one of them, but
it appears that the loss of one degree of freedom makes the chain very
sticky, that is, slow to converge. As an alternative, we assign
zero-centered normal prior
distributions to all $\psi_e$, and use a transformation group move
[\citet{refLiuWu99JASA}; \citet{refLiuSabatti00Biometrika}; \citet{refLiu01book}] to improve mixing of the MCMC sampler. Specifically,
we first propose a move by adding a random number $z$ to all $\psi_e$
and subtracting $z$ from all $\phi_g$, and then use the
Metropolis--Hastings  rule  to accept or reject this move.  Since we
only care about the relative phases of genes and
experiments, we use $\phi_g+\psi_1$ as the gene's relative phase and
$\psi_e-\psi_1$ as
the phase for an experiment.

 \item When a gene violates the assumption that its peaking time in
 the cell cycle  relative to all other genes is fixed  across
 different experiments,
% the assumption of fixed relative peaking time among genes is not
%true for a gene compared to all other gene,
its multiple time
series will show inconsistent phases, which leads to multiple
modes for its phase parameter $\phi_g$ and amplitude parameters
$A_{ge}$. It is difficult to get out of this kind of local mode by
updating $\phi_g$ and $A_{ge}$ separately and locally. We combine
the idea of grouping [\citet{refLiuKong94Biometrika}] and
Metropolized independence sampling
[\citet{refHastings70Biometrika}; \citeauthor{refLiu01book} (\citeyear{refLiu96StatComput,refLiu01book})]
to deal with this kind of local mode. We call it the
Metropolized independence group sampler (MIPS). We
first propose a new $\phi_g$ independent of old $\phi_g$, say,
from its prior distribution or an approximation of its conditional
posterior distribution. Then, we sample all $A_{ge}$ conditional on
the new $\phi_g$. The Metropolis--Hastings rule is used to decide
whether to accept this move or not. To get a good proposal of
$A_{ge}$, we use linear regression to get the least square
estimate of $A_{ge}$ and use it as the center of the proposal
distribution of $A_{ge}$.

 \item We again use MIPS to deal with the strong correlation within
   the trend parameters ($a_{ge}, b_{ge}, c_{ge},
d_{ge}$) for a time series. The key is to propose a new $d_{ge}$
independent of the old $d_{ge}$ and sample ($a_{ge}, b_{ge}, c_{ge}$)
jointly conditional on the new $d_{ge}$, which is a multivariate
normal distribution here.

 \item There are also strong correlations between $\lambda_e$ and all
   $A_{ge}$ of
the same experiment $e$. We still use MIPS to perturb the MCMC
chain. We propose a new $\lambda_e$ independent of the old $\lambda_e$
and sample all $A_{ge}$ of the same experiment $e$ conditional on
the new $\lambda_e$. Similar to the MIPS moves for $\phi_g$ and
$A_{ge}$ of the same gene $g$, we used the least square estimate of
$A_{ge}$ to improve the proposal efficiency.

\end{itemize}

It should be noted that MIPS improves the mixing of the MCMC
chain, especially at the initial state of the sampling, with an
extra cost in computation. Our simulations indicated that this is
a worthy effort. In meta-analysis, it is not unusual that
different experiments support different values for a shared
parameter. As a result, the shared parameter may have a
multi-modal distribution. In that case, strategies such as MIPS
for making global moves are desirable.
\end{appendix}

\section*{Acknowledgments}

We  thank Professor Xiao-Li Meng for his helpful
suggestions that led to a modification of the model in this paper.
We are also grateful to Professor Wing H. Wong and Dr. Xin Lu for
their valuable comments.
%This research is supported in part by the
%NIH grant R01GM078990 and the NSF grant DMS-0706989.
All R codes and fitting results are available upon request.

\begin{supplement}[id=suppA]
\stitle{Various supporting materials}
\slink[doi]{10.1214/09-AOAS300SUPP}
\slink[url]{http://lib.stat.cmu.edu/aoas/300/supplement.pdf}
\sdatatype{.pdf}
\sdescription{In this supplement we provide
model fitting diagnoses, hierarchical clustering results, the
effect of data size on the statistical power, supporting evidences
for newly found genes, and figures referred to in this paper.}
\end{supplement}

\printaddresses

\end{document}